\documentclass[reprint,amsmath,amssymb,aip,apl]{revtex4-2}

\usepackage{graphicx}
\usepackage{dcolumn}
\usepackage{bm}
\usepackage[utf8]{inputenc}
\usepackage[T1]{fontenc}
\usepackage{mathptmx}
\usepackage{etoolbox}
\usepackage{xcolor}
\usepackage{overpic}
\begin{document}

\preprint{AIP/123-QED}

\title{Real-Time FPGA-Based Multi-Parameter Feedback Stabilization of a Silicon Double Quantum Dot}

\author{Johnathan Bryan}
\email[Corresponding author: ]{jbryan@physics.ucla.edu}
\affiliation{Department of Physics and Astronomy, University of California, Los Angeles, California 90095, USA}

\author{Tim J. Wilson}
\affiliation{Department of Physics and Astronomy, University of California, Los Angeles, California 90095, USA}
\affiliation{National Institute of Standards and Technology, Gaithersburg, Maryland 20899, USA}

\author{Hong-Wen Jiang}
\affiliation{Department of Physics and Astronomy, University of California, Los Angeles, California 90095, USA}

\date{\today}

\begin{abstract}
Long term operation of semiconductor spin qubits requires active stabilization of quantum dot potentials against low-frequency charge noise. To address this, we demonstrate a gradient-descent feedback approach in a silicon double quantum dot device using high-bandwidth radio-frequency (rf) reflectometry readout and a field-programmable gate array (FPGA) for digital signal processing. This architecture enables continuous multi-parameter gradient calculation and quick gate voltage updates, significantly increasing the effective feedback bandwidth compared to previous methods relying on slow direct-current transport measurements. By operating with 8 steps per feedback cycle and an integration time of 25.6 $\mu$s, this high-speed stabilization scheme maintains better than -6 dB of noise suppression up to 5 kHz, achieving an attenuation of -7.4 ± 0.2 dB at the maximum measured frequency. This effectively suppresses low-frequency 1/f noise, maintaining device stability over longer time periods, and potentially enables real-time control in large-scale quantum dot arrays.
\end{abstract}

\maketitle

\section{Introduction}

Semiconductor spin qubits are a promising platform for fault-tolerant quantum computation due to their potential for high-fidelity operation, small footprint, and compatibility with modern semiconductor manufacturing \cite{zhang2018, burkard2023, veldhorst2014, loss1998, hanson2007, zwanenburg2013, vandersypen2017, maurand2016, xue2022}. The realization of these large-scale architectures, however, requires precise control over the local electrostatic environment. A primary decoherence mechanism in these systems is low-frequency $1/f$ charge noise, which arises from fluctuating charge traps in the surrounding device materials and limits free-evolution dephasing times \cite{yoneda2017, wilson2026, dial2013, eng2015, muhonen2014, paladino2014, connors2022, connors2019}. Over time, this noise alters the local potential landscape, causing the operating point of quantum dot systems to drift and degrading long-term qubit operation and $T_{2}$ coherence times \cite{vepsalainen2022, petit2018, freeman2016, struck2020}. 

Active feedback stabilization offers a pathway to mitigate slow fluctuations \cite{yang2011, nakajima2021, ye2024}. For instance, Nakajima \textit{et al.} utilized proportional-integral-derivative (PID) feedback using a field-programmable gate array (FPGA) to maintain a constant current through a sensing dot, achieving a feedback bandwidth of $\sim100$ kHz \cite{nakajima2021, nakajima2020}. While demonstrating high-bandwidth charge-sensor stabilization, maintaining a target DQD potential configuration requires multi-parameter control. As devices scale, multi-parameter control algorithms will become necessary for larger arrays of quantum dots to implement feedback stabilization. More recently, Wen \textit{et al.} successfully applied a multi-dimensional gradient descent algorithm to a silicon DQD to maintain a target potential configuration without becoming trapped in constant-current loops, measuring a feedback bandwidth of up to 300 mHz \cite{wen2025}. However, relying on standard DC transport current measurements limited their effective suppression capability.

\begin{figure}[h]
    \centering
    \includegraphics[width=0.45\textwidth]{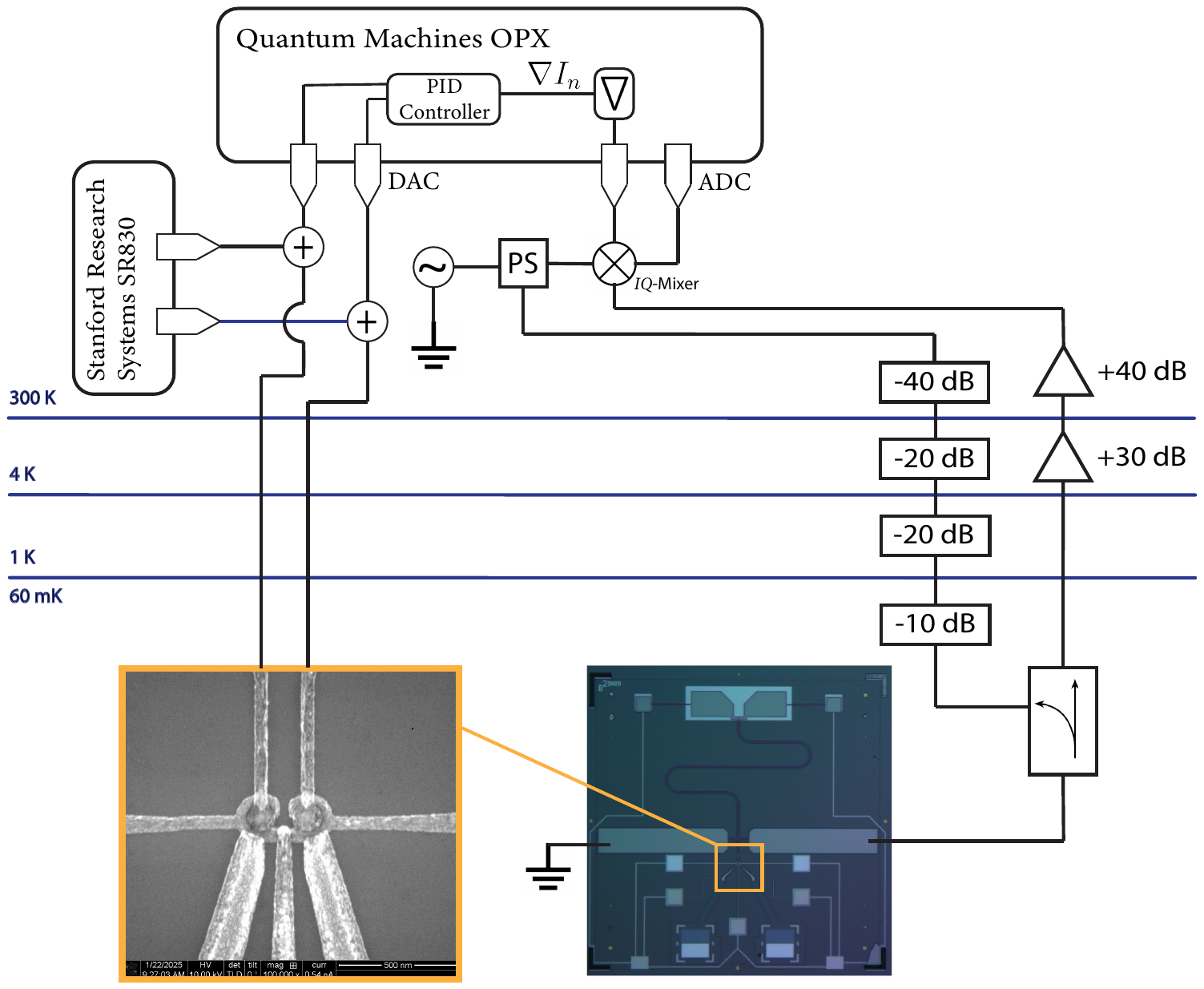}
    \caption{Schematic of the experimental setup. The double quantum dot, located at the 60 mK stage, is coupled to an rf-reflectometry circuit for fast dispersive readout. The reflected signal is amplified and demodulated, while a Quantum Machines OPX provides digital signal processing and voltage control.}
    \label{fig:setup}
\end{figure}

In this work, we build upon these approaches by implementing a high-bandwidth, two-axis gradient descent feedback algorithm using fast radio-frequency (rf) reflectometry for dispersive readout \cite{reilly2007, petersson2010, vigneau2023, ares2016, zheng2019, noiri2020, schaal2020}. A schematic of the measurement circuit and control architecture is shown in Fig.~\ref{fig:setup}.

By coupling the DQD to a microwave resonator and using the OPX by Quantum Machines for digital signal processing and voltage control, we implement an active stabilization scheme to reduce low frequency 1/f noise in the system. This approach increases the measurement and feedback bandwidth, yielding a better than $-6$ dB noise suppression up to 5 kHz.

\section{Experimental Setup}

\begin{figure*}[t]
    \centering
    \begin{overpic}[width=0.48\textwidth]{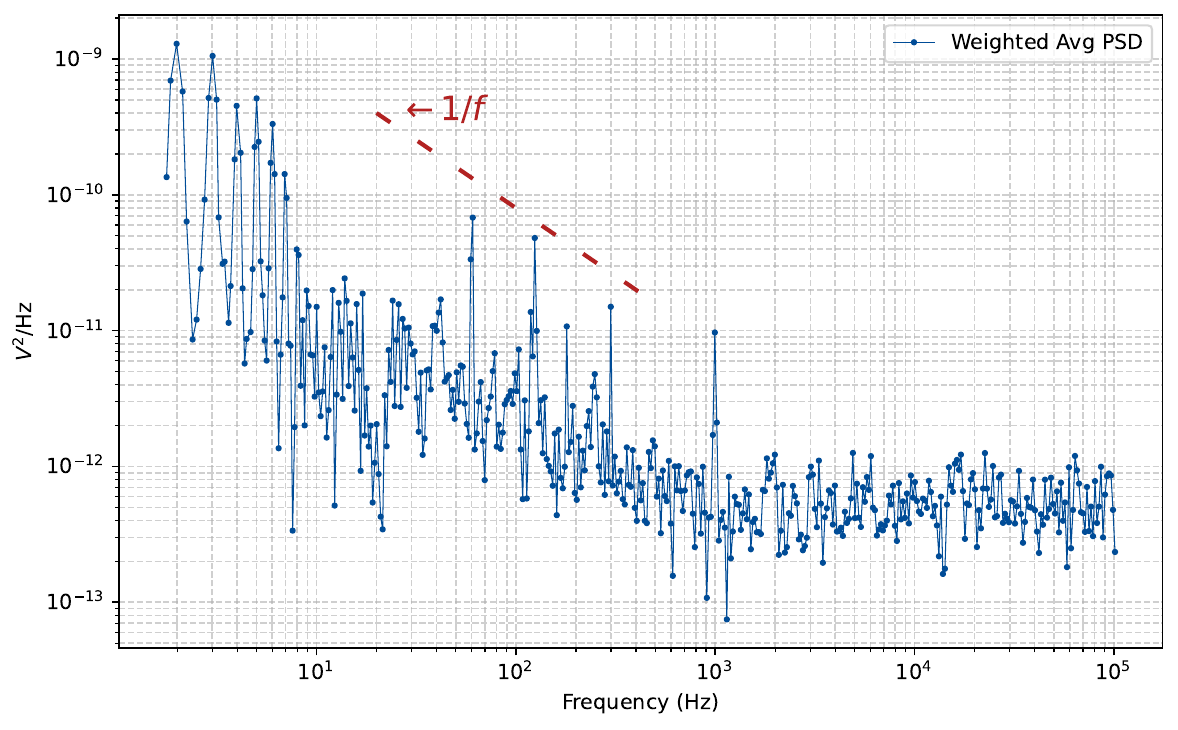}
        \put(-5,55){\large \textbf{(a)}}
    \end{overpic}
    \hfill 
    \begin{overpic}[width=0.48\textwidth]{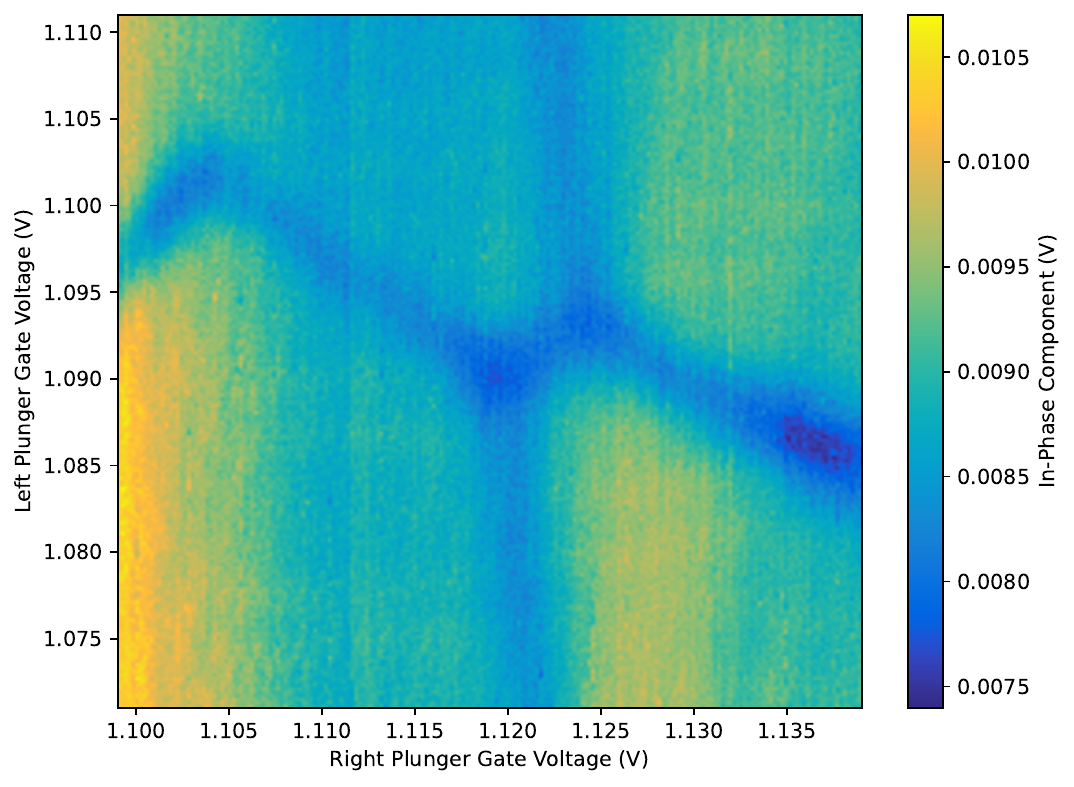}
        \put(-5,67){\large \textbf{(b)}}
    \end{overpic}
    
    \caption{(a) Power spectral density of the in-phase signal component showing the $1/f$ noise present in our system, measured using an SR785 signal analyzer. A dashed reference line indicates the slope corresponding to pure 1/f noise. This spectrum represents a weighted average of multiple measurement runs with varying spans, recorded at the operating point located at the center of (b). (b) Device charge stability diagram. An SR830 provides a DC bias offset to both plunger gates, while the OPX is responsible for sweeping the gate voltages and providing the active feedback voltage. Feedback was initialized at a minimum in the interdot transition signal at the center of the diagram.}
    \label{fig:psd_and_stability}
\end{figure*}

The device measured in this experiment is a silicon DQD. The underlying heterostructure consists of a 225 nm Si$_{0.7}$Ge$_{0.3}$ buffer, a 5 nm Si quantum well, and a 50 nm Si$_{0.7}$Ge$_{0.3}$ cap. Phosphorus-doped regions provide ohmic contacts to the two-dimensional electron gas (2DEG) source and drain channels. The overlapping gate architecture consisting of two Al metallic gate layers where the plunger gates are isolated from barrier gates by an oxygen plasma enhanced native alumina. The gates are patterned atop an atomic layer deposition (ALD) grown aluminum oxide layer with thicknesses of 5 nm and 25 nm. The device used is the same one characterized by Wilson \textit{et al.} \cite{wilson2026}. All measurements were taken at a base temperature of 60 mK.

To enable fast readout and quick integration times, the setup utilizes rf-reflectometry. The reflectometry circuit incorporates an rf source and an IQ mixer, with the reflected signal passing through an amplification chain consisting of a +30 dB amplifier at the 4 K stage and a +40 dB amplifier at room temperature (300 K). At optimal tuning, this setup achieves a signal-to-noise ratio (SNR) of unity at an integration time of 34.54 ns \cite{wilson2026}. 

Control and stabilization of the device are managed by a combination of DC and high-speed digital electronics. A Stanford Research Systems SR830 is used to supply the necessary DC bias offsets. High-speed control is implemented with a Quantum Machines OPX, executing the gradient-based feedback algorithm on its FPGA, calculating gradient estimates and updating the gate voltages through integral feedback as indicated by Equation \ref{eq:GD}.

Before turning on the feedback, we map the device's charge stability diagram and start the feedback loop at a minimum in the interdot transition signal. At this starting point, the device shows the typical $1/f$ charge noise that our setup is designed to suppress, as shown in Fig.~\ref{fig:psd_and_stability}. The feedback is initialized at a local minimum of the interdot transition signal, providing a well-defined operating point for the stabilization algorithm.

\section{Gradient Descent Feedback Protocol}

To stabilize the system against environmental fluctuations, we employ an alternating-axis gradient descent technique executed on the OPX's FPGA. The state of the DQD is read out by probing the coupled CPS resonator. Variations in the local electrostatic environment alter the quantum dot potential landscape, which in turn manifest as measurable shifts in the in-phase ($I$) component of the reflected microwave signal.

The OPX counteracts this charge noise by calculating a local gradient and adjusting the feedback output voltages. The protocol operates through a multi-parameter feedback algorithm, where a single iteration proceeds in the following steps:

First, a baseline measurement of the in-phase component is recorded; this baseline measurement is shared for all gradient component calculations in the current cycle. The voltage on the $i$-th gate is then stepped by a small value $\Delta V$ (set to $10$ mV in this experiment). Following this step, the in-phase component is remeasured. The voltage is subsequently returned to its original value. This measurement and perturbation sequence is repeated for each gate included in the stabilization loop, in our case, the left and right plunger gates of our device.

Once the differential measurements are collected for all active gates, the local gradient for each axis is defined simply as the difference between the stepped and baseline signals, $\nabla I_{N,i} = I_{\text{stepped},i} - I_{\text{baseline}}$ (Note that because $\Delta V$ is constant, the division step is omitted and implicitly absorbed into $K_i$ and $K_p$). Since the feedback loop actively seeks to drive this gradient to zero, the measured gradient $\nabla I_{N,i}$ serves directly as the error signal. Here, we define $V_{N,i}$ and $V_{N+1,i}$ as the applied voltages on the $i$-th gate at the current step $N$ and the subsequent step $N+1$, respectively. Additionally, $K_p$ and $K_i$ represent the proportional and integral gain coefficients. The output voltages are then updated according to the incremental PID equation \cite{astrom1997}:\begin{equation}V_{N+1,i} = V_{N,i} + K_{p}\nabla I_{N,i} + K_{i}\sum_{N'} \nabla I_{N',i}\label{eq:GD}\end{equation}

For this experiment, we set the proportional gain coefficient $K_p$ to zero. Since the calculated error signal relies on the numerical derivative of a noisy measurement, proportional control is susceptible to high-frequency noise present in the system. During the feedback protocol, the algorithm applies a $1~\mu$s wait time after stepping the voltage to ensure a stable output, followed immediately by a $25.6~\mu$s measurement time. Accounting for all baseline and stepped measurements across both gates, a single complete feedback cycle executes in $81.8~\mu$s. By continuously cycling through this protocol, the local dot potentials remain at their target configuration, effectively suppressing the $1/f$ noise present in our system.

\section{Results and Discussion}

\begin{figure}[h]
    \centering
    \begin{overpic}[width=0.45\textwidth]{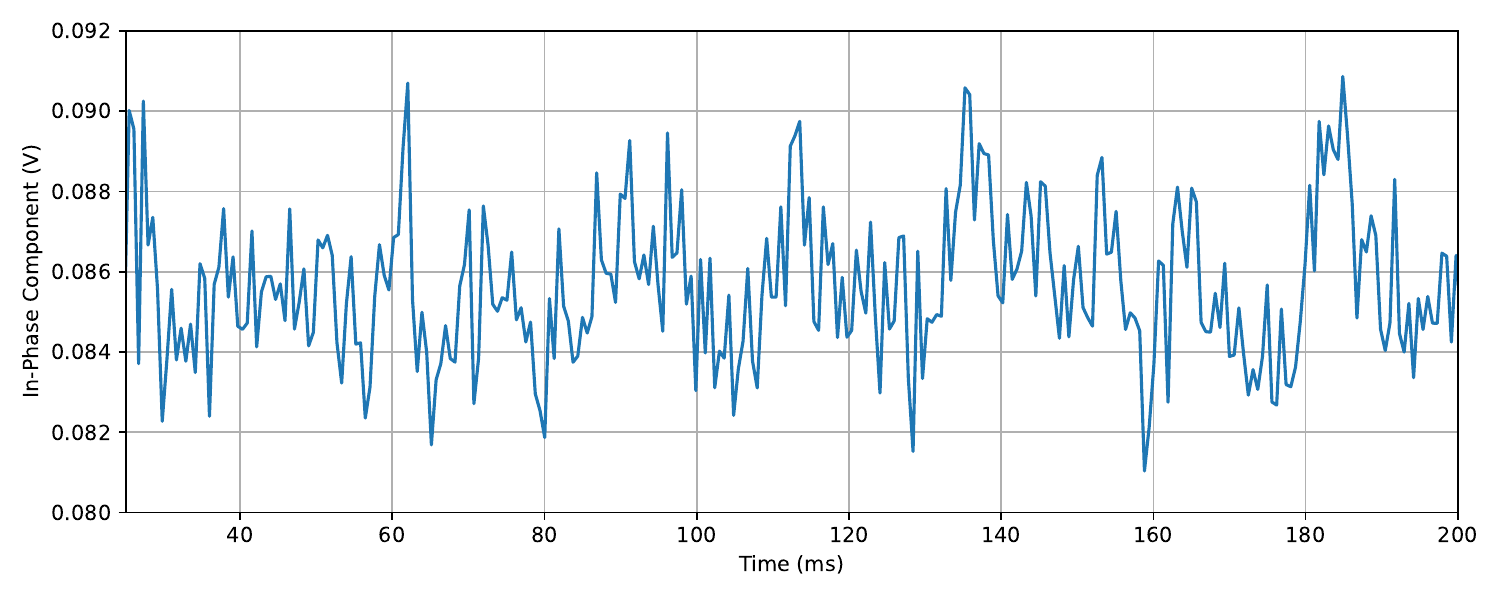}

        \put(-5,35){\large \textbf{(a)}}
    \end{overpic}
    
    \vspace{0.2cm} 
    
    \begin{overpic}[width=0.45\textwidth]{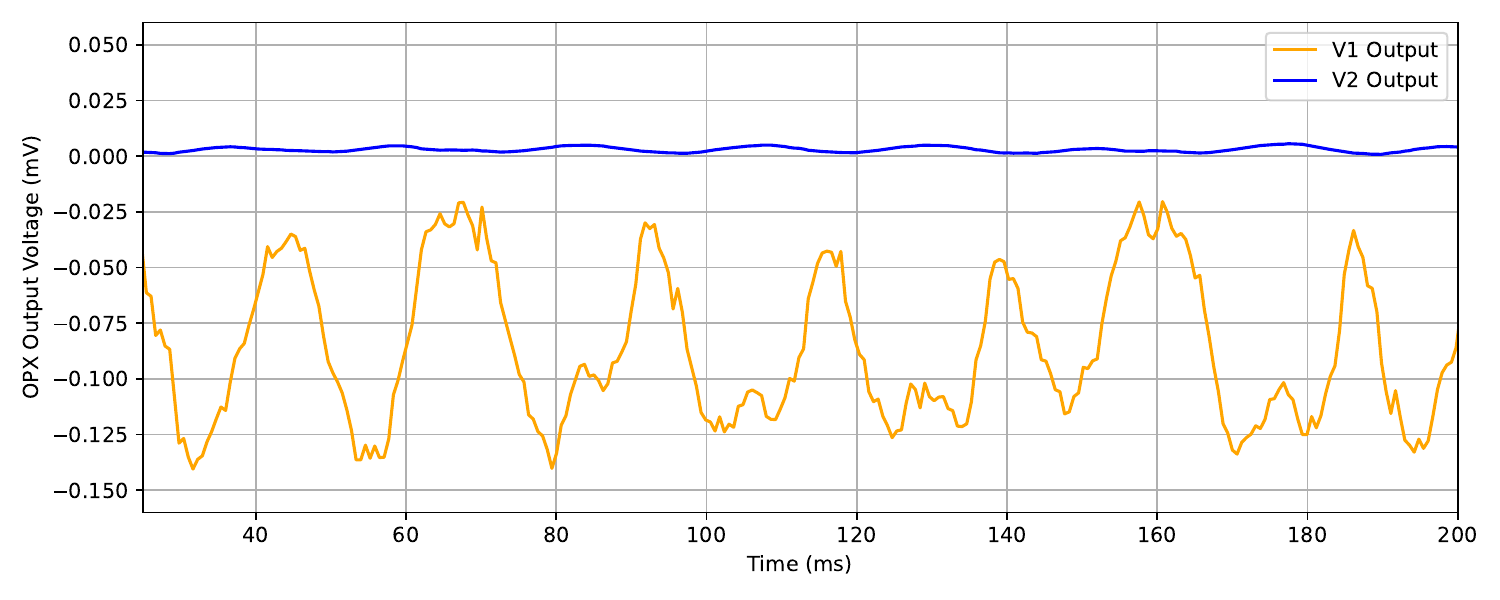}
        \put(-5,35){\large \textbf{(b)}}
    \end{overpic}
    
    \caption{Time-domain response of the feedback protocol. (a) shows the in-phase signal over time as recorded by the OPX. A sinusoidal perturbation is artificially introduced to evaluate the algorithm's tracking performance. (b) shows the output voltages, $V_1$ and $V_2$, applied by the OPX to the two plunger gates to dynamically compensate for the disturbance. The oscillatory response of \(V_1\) is consistent with compensation of the applied perturbation. A small output in $V_2$ can also be seen.}
    \label{fig:time_domain}
\end{figure}

To observe the algorithm's dynamic response, we input a low-frequency sine wave onto one of the device's plunger gates. By directly recording the signal received by the OPX alongside the compensating output voltages ($V_1$ and $V_2$) as a function of time, we can visualize the feedback loop actively counteracting this disturbance (Fig.~\ref{fig:time_domain}).

\begin{figure}[h]
    \centering
    \includegraphics[width=0.45\textwidth]{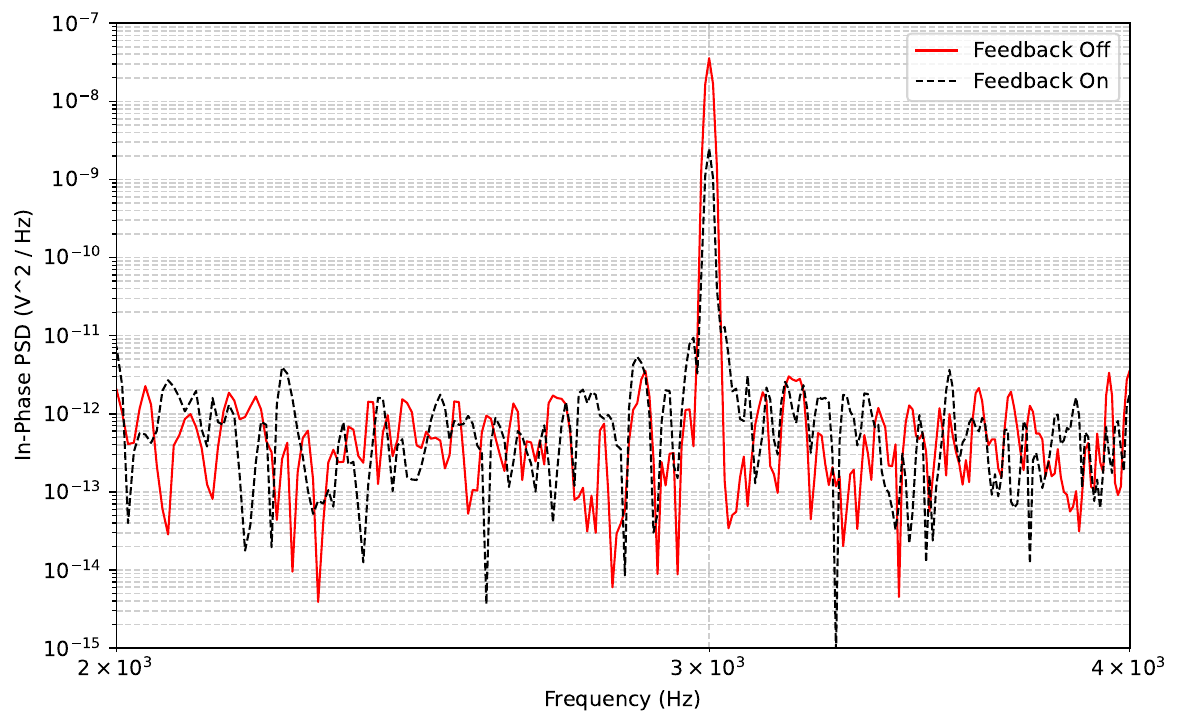}
    \caption{Power spectral density of the in-phase component with a 3 kHz sinusoidal perturbation applied to the device. The solid red line represents the spectrum with the feedback loop disabled, while the dashed black line is with feedback enabled.}
    \label{fig:psd}
\end{figure}

We then evaluate the algorithm's ability to suppress this perturbation by measuring the power spectral density (PSD) of the in-phase component of the signal with feedback on and with feedback off. As shown in Fig.~\ref{fig:psd}, when introducing a simulated 3 kHz perturbation, the feedback protocol achieves over -10 dB of suppression. For these bandwidth measurements, we optimized $K_i$ independently for each perturbation frequency. We did this by increasing $K_i$ until we no longer saw any increase in measured attenuation in the PSD of the in-phase signal.

\begin{figure}[h]
    \centering
    \includegraphics[width=0.45\textwidth]{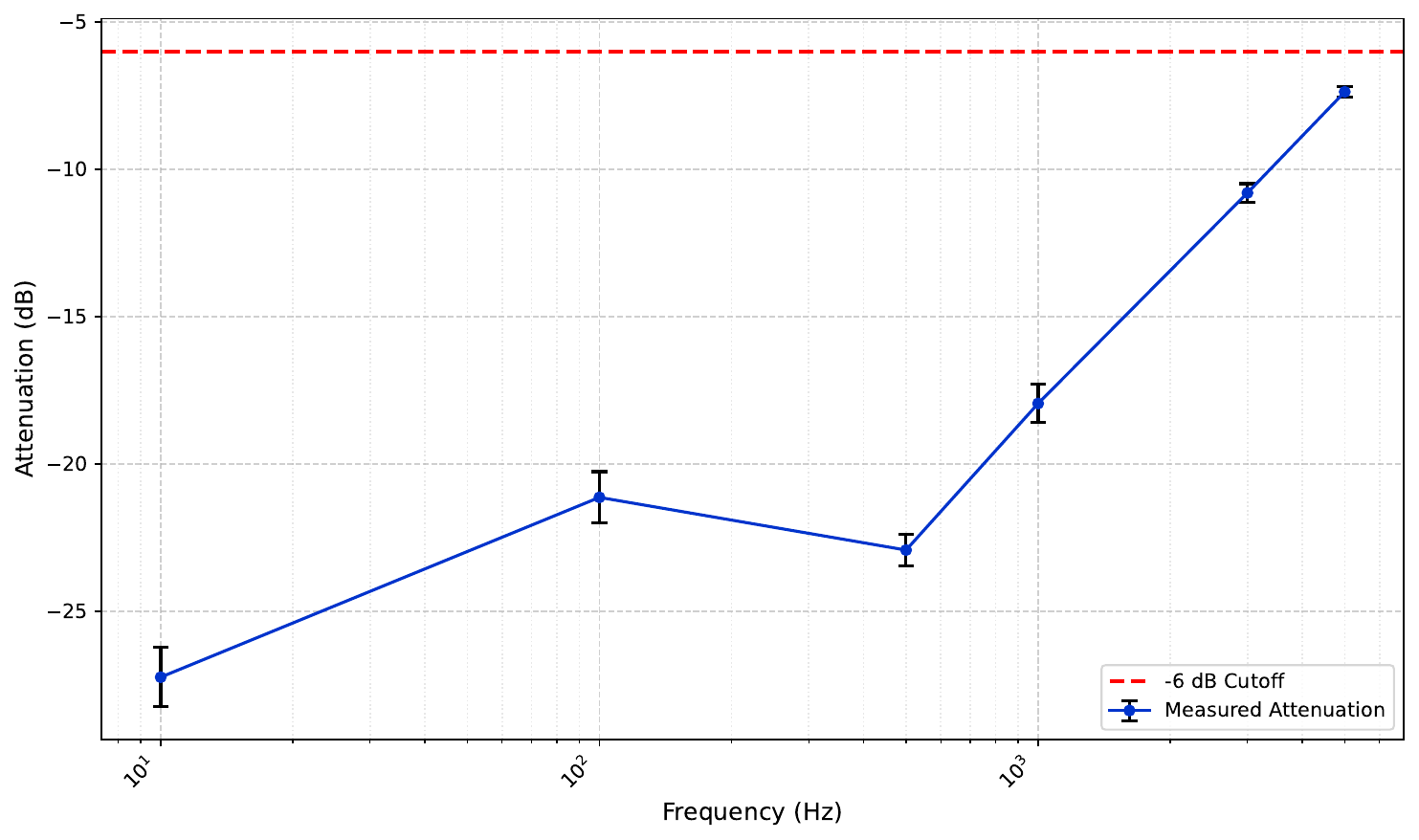}
    \caption{Measured attenuation of the in-phase signal power spectral density as a function of the applied perturbation frequency. The attenuation of the highest frequency data point shown is -7.4 ± 0.2 dB. Error bars represent the standard error of the measurement}
    \label{fig:bandwidth}
\end{figure}

To characterize the performance of the feedback algorithm, we input different frequency sine waves onto one of the plunger gates and measured the attenuation in the PSD, observing the largest attenuation at low frequencies and maintaining better than -6 dB of suppression up to 5 kHz, with the highest frequency data point showing an attenuation of -7.4 ± 0.2 dB (Fig.\ref{fig:bandwidth}). This represents a substantial improvement over previous multi-parameter implementations relying on DC transport measurements, which achieved similar attenuation at 0.3 Hz.

While the present experiment focuses on stabilizing the potential landscape of a DQD, applying feedback updates during coherent qubit operations could perturb the qubit Hamiltonian and degrade coherence. Consequently, in order to utilize this algorithm, feedback would therefore need to be coordinated with qubit operation, for example by interleaving stabilization and computation.

The maximum effective feedback bandwidth is currently limited by the total duration of our feedback cycle, which is set by our long integration times. Utilizing a 25.6 $\mu$s integration time and 1 $\mu$s voltage transition times, our feedback protocol requires 81.8 $\mu$s per cycle. According to the Nyquist-Shannon sampling theorem, the maximum cutoff frequency for this cycle time is approximately 6.1 kHz \cite{shannon1949}. Because our 5 kHz bandwidth closely approaches this limit, further improvements require reducing the feedback cycle time by increasing the measurement SNR. 

\section{Conclusion}

In summary, we have demonstrated a high-bandwidth, multi-parameter feedback stabilization protocol for a silicon double quantum dot. By integrating fast readout via rf-reflectometry with an FPGA-based Quantum Machines OPX controller, we demonstrated suppression of low-frequency electrostatic disturbances using active feedback. The system maintained better than -6 dB noise attenuation up to 5 kHz, achieving -7.4 ± 0.2 dB at the maximum measured frequency. This extends the demonstrated noise-suppression range to 5 kHz, compared with approximately 0.3 Hz in previous multidimensional feedback implementations using DC transport measurements \cite{wen2025}. Because reducing low-frequency fluctuations is a critical step toward reliable quantum computation, future work will focus on characterizing the direct impact of this interleaved feedback loop on spin qubit coherence times and gate errors, paving the way for real-time control in scalable quantum dot arrays \cite{vepsalainen2022}.

\section*{Author Declarations}
\subsection*{Conflict of Interest}
The authors have no conflicts to disclose.

\section*{Data Availability}
The data that support the findings of this study are available from the corresponding author upon reasonable request.

\begin{acknowledgments}
This research was sponsored by the Army Research Office (ARO) and was accomplished under Grant No. W911NF-23-1-0016 at UCLA.
\end{acknowledgments}

\subsection*{Notes}
J. B. performed the measurements, analyzed the data, and prepared the original manuscript. T. J. W. fabricated the device and contributed to writing and revising the manuscript. H.-W. J. supervised the project. All authors discussed the results and contributed to the final manuscript.

\bibliography{references}

\end{document}